\documentclass[12pt]{article}

\usepackage{amssymb,amsmath}
\usepackage{url}

\usepackage{graphicx}

\newcommand{\be}{\begin{equation}}
\newcommand{\ee}{\end{equation}}

\newcommand{\dlt}{\delta}

\newcommand{\bt}{\beta}

\newcommand{\al}{\alpha}
\newcommand{\ra}{\rightarrow}

\newcommand{\gm}{\gamma}
\newcommand{\om}{\omega}
\newcommand{\Om}{\Omega}

\newcommand{\lbd}{\lambda}

\newcommand{\cL}{{\cal L}}
\newcommand{\bE}{{\bf E}}
\newcommand{\bH}{{\bf H}}
\newcommand{\br}{{\bf r}}
\newcommand{\rgl}{\rangle}
\newcommand{\lgl}{\langle}

\begin{document}

\begin{center}

{\Large{\bf Self-organization in complex systems as decision making} \\[5mm]

V.I. Yukalov$^{1,2*}$ and D. Sornette$^{1,3}$} \\ [3mm]

{\it
$^1$Department of Management, Technology and Economics, \\
ETH Z\"urich, Swiss Federal Institute of Technology, \\
Scheuchzerstrasse 7, Z\"urich CH-8092, Switzerland \\ [3mm]

$^2$Bogolubov Laboratory of Theoretical Physics, \\
Joint Institute for Nuclear Research, Dubna 141980, Russia \\ [3mm]

$^3$Swiss Finance Institute, c/o University of Geneva, \\
40 blvd. Du Pont d'Arve, CH 1211 Geneva 4, Switzerland}

\end{center}

\vskip 2cm

\begin{abstract}

The idea is advanced that self-organization in complex systems can be 
treated as decision making (as it is performed by humans) and, vice versa, 
decision making is nothing but a kind of self-organization in the decision 
maker nervous systems. A mathematical formulation is suggested based on 
the definition of probabilities of system states, whose particular cases 
characterize the probabilities of structures, patterns, scenarios, or 
prospects. In this general framework, it is shown that the mathematical 
structures of self-organization and of decision making are identical. This 
makes it clear how self-organization can be seen as an endogenous decision 
making process and, reciprocally, decision making occurs via an endogenous 
self-organization. The approach is illustrated by phase transitions in 
large statistical systems, crossovers in small statistical systems, 
evolutions and revolutions in social and biological systems, structural 
self-organization in dynamical systems, and by the probabilistic formulation 
of classical and behavioral decision theories. In all these cases, 
self-organization is described as the process of evaluating the probabilities 
of macroscopic states or prospects in the search for a state with the 
largest probability. The general way of deriving the probability measure 
for classical systems is the principle of minimal information, that is, the 
conditional entropy maximization under given constraints. Behavioral biases
of decision makers can be characterized in the same way as analogous to quantum fluctuations 
in natural systems.   

\end{abstract}

\vskip 1cm


\vskip 3cm

$^*$Author for correspondence (yukalov@theor.jinr.ru)

\newpage

\section{Introduction}

In order to avoid misunderstanding, we would like to stress from the very 
beginning that the main idea and the principal novelty of the present article
is the demonstration that {\it self-organization of complex systems can be 
described in precisely the same mathematical terms as decision making by 
human beings}. In that sense, these two, to the first glance, absolutely 
different processes, can be understood as different representations of the
same phenomenon. To realize this demonstration, we resort to the optimization 
methods based on conditional maximization of entropy, or minimization of 
an information functional. However, one has to keep in mind that the 
optimization methods, as such, are not the goal but are the {\it tools}
for accomplishing the demonstration of the equivalence of self-organization 
and decision making. 

Before plunging into the mathematical formulation, we give in this introduction 
below the general feeling of the problem, which should help the reader to grasp 
the main ideas that will be mathematically proved in the following sections. 
      
The universe is marvelously structured. Everywhere and at any scale one 
examines, one cannot escape a deep sense of wonder about the origin and 
meaning of the remarkable organizations that can be observed, exhibiting 
complex interplays between regularity and irregularity, order and disorder, 
periodicity and stochasticity, aesthetics and randomness. Confronted with 
these impressions, a first reaction is to invoke the presence of a 
superior being, whose will has determined the ``natural order of things'' 
and who is in charge with its maintenance. In the standard teleological 
or ``intelligent design'' argument (McPherson 1972), it is pointed out 
that the delicate and harmonious work of a clock requires the expert 
agency of a watchmaker. Then, by analogy, the organization of the universe
mentioned above cannot be conceived without the existence, will and agency 
of a super-watchmaker. This argument that design implies a designer 
(Aquinas 13th century, Paley, 1802) is permeating in one or another form
all types of religious beliefs that have appeared over the history of 
humanity, perhaps as far as 200 000 years ago (Dunbar 2006), in the search 
for meanings and the quest for permanence. 

Many thinkers and scientists have contributed to the rebuttal of the 
intelligent design argument. On pure logical grounds, the proposition
that a designer creates by definition a design, which has some structure, 
does not reverse logically into the proposition that a structure is necessary 
by design, and thus requires a designer. Indeed, this would amount
to assuming incorrectly the equivalence of  $A \to B$ and $B \to A$.
Here $A$ is the proposition of the existence of designer and $B$ is
the property of a system to possess structure. The incarnation of the fact 
that, if $A \to B$ holds true, this does not imply the validity of $B \to A$, 
is found in the concept  of self-organization and of emergence, namely that 
novel organized behaviors emerge from spontaneous collective organization. The 
scientific theory of self-organization has matured in science over most of the 
20th century in different disciplines. The ubiquity of self-organization has 
made irrelevant (or better said, unnecessary and non-parsimonious) the concept 
of a top-down design (and of control) by a super-being (Kauffman 1996). In 
contrast, we now understand that spontaneous bottom-up self-organizations occur 
generically and provide the mechanisms and the construction processes for 
explaining most, if not all, phenomena of the Universe. Self-organization can be 
defined as the spontaneous formation in a complex system of global structures out 
of local interactions (Haken 2005; Heylighen 2009). By global structures is meant 
spatial, temporal, or functional structures involving the system as a whole. 
Examples of such ubiquitous structures has been the objects of numerous studies 
in the physical and biological sciences (Glansdorff and Prigogin 1971; Haken 1983; 
Nicolis 1986) and in cybernetics (Ashby 1947; von Foerster 1960, 1999; 
von Foerster and Pask 1960; Wiener 1961; von Glasserfeld 1996; Labini et al. 2009)
as well as, to a lesser degree, in social sciences (Schelling 1978; Krugman 1996; 
Brock and Hommes 1997; Galam 2012). More recently, great attention is being paid 
to the processes of self-organization in various networks, including neuron networks 
in the brain (Mainzer 2007; Chialvo 2010; Werner 2012; Fingelkurts et al. 2012)
and of self-organization of various swarms and flocks 
(Vanni et al. 2011; Turalska et al 2011), where self-organization is understood 
as the unexpected appearance of collective or coherent behavior that is termed 
{\it swarm intelligence}. Complex behaviors of simple physical systems, imitating
a kind of trivial intelligence, can be due to entropic forces 
(Wissner-Gross and Freer 2013). 

The goal of the present article is to suggest a novel level of understanding,
combining self-organization of the complex system and decision making of that 
same complex system (without invoking the will and decision making of any 
controller or watchmaker). This is based on the recognition that the 
{\it mathematical structures} of self-organization and of decision making are 
identical. In other words, the process of self-organization corresponds to an 
endogenous decision making process, giving the impression that a superior intention 
is at work. While this view point has been made many times by philosophers of the 
science of complexity, for instance, by Hooker et al. (2011) (and references therein), 
we provide, what we think, is the first strict {\it mathematical formulation} of 
this equivalence. By recognizing the endogenous nature of decision making that is 
embedded in any self-organization process, we clarify the meaning of intention 
and will, which are possessed by the complex system itself. While this has 
important philosophical implications, the main characteristic of our approach
is a rigorous {\it mathematical framework}, whose precise language permits the 
demonstration of the proposed equivalence.

As any endeavor touching such big concepts and existential questions, there 
are many roots and precursors of our ideas, going back to Plato and Aristotle, 
Saint Anselm around 1000 AD and Kant. Perhaps the idea closest to our 
proposition is the one formulated by the German philosopher Kant. In 1790, 
in his Kritik der Urteilskraft, whose translation can be found in (Kant, 2007), 
he introduced the term ``self-organization". He argued that it is possible 
for an entity to exist, whose parts or organs are simultaneously ends and means. 
Such a system of organs must be able to behave as if it has a mind of its own, 
that is, it is capable of governing itself. This idea can be understood as if 
a system could decide on the process of self-organization. In other words, 
understood from the theme we propose here, self-organization can be interpreted 
as the process of decision making performed by a complex system. And 
reciprocally, decision making can be treated as the process of self-organization 
of the information by humans in the process of knowledge gathering  
(Poerksen 2003; von Foerster 2003). The idea of the similarity between statistical 
systems and game-theory forms (Galam and Walliser 2010) is a particular example 
of the approach we follow in the present article.

We prove the general proposition of the similarity between self-organization and 
decision making by developing an explicit {\it mathematical formulation} 
describing these processes in the frame of the same general {\it probabilistic} 
approach. Necessarily, such an approach has to be probabilistic for two reasons. 
First, a probabilistic approach is mathematically more general, including the 
deterministic one as a particular case. Second, observed processes in nature 
are practically always stochastic, and therefore require a probabilistic 
description. A system can often be in several macroscopic states, or several 
different structures can be formed. Which of the states is mostly occupied, 
or which structure prevails, is defined by the corresponding probability 
distributions, which in full generality are also path or history dependent. 
Although, under the same conditions, one of the structures, or patterns, can 
be preferable and be more often realized, other types of structures can also 
occur, though with a smaller probability or weaker frequency.

Armed with the notion that the description of any system requires a 
probabilistic framework, we can now give a first intuition of the 
correspondence between the processes of self-organization and decision making, 
which can be described in the frame of the same mathematical framework with 
just a slight change of terminology. Respectively, both processes can be 
defined in similar words:

\vskip 2mm
\noindent
{\it Self-organization is the process of evaluating the probabilities of 
system states in the search for the most stable state}.

\vskip 1mm
\noindent
{\it Decision making is the process of evaluating the probabilities of 
decision prospects in the search for the most preferable prospect}.
    
\vskip 2mm

We start Sec. 2 by explaining the intimate connection between the notions
of stability and of probability, so that the system state is more probable, provided it is
more stable. Then we give the general scheme outlining the analogies between 
the system states and decision prospects in the frame of the probabilistic 
representation. The principle of minimal information, described in Sec. 3,
serves as a basic tool for defining the explicit forms of the state or
prospect probabilities. Section 4 illustrates the general approach by 
particular examples of many-body, or multi-agent, systems, including 
large and small statistical assemblies, systems with mesoscopic fluctuations,   
and social and biological systems. The dynamics of self-organization in
strongly out-of-equilibrium systems is described in Sec. 5, where the 
probabilities of structure formation are shown to be defined by the 
system expansion exponents. An illustration of the use of the expansion
exponents for the description of the structures arising under turbulent 
photon filamentation is given. The reformulation of decision making in 
probabilistic terms, which makes it analogous to self-organization, 
is presented in Sec. 6 for both classical and quantum variants. 
Section 7 concludes.

We would like to stress that our main aim is not to analyze in detail the 
behavior of particular systems, but to show that their final mathematical 
description can be realized in the same general {\it probabilistic framework}. 
Particular examples that we mention for illustration can be already well 
understood. This does not impact the fact that the principal novelty of our 
article is the demonstration in mathematical terms that the general features 
of self-organization in any complex system are formally equivalent to the 
process of decision making by alive beings.

\section{System states as decision prospects}

The possibility of characterizing (i) the states of empirical systems as
decision prospects and (ii) transitions between the states as if the system
would be deliberating choosing the most stable, that is, the most preferable 
state, relies on the fact that observation and knowledge acquisition always 
require the existence of an observer. In the process of describing and 
characterizing the system of interest, the observer ascribes to the system 
a kind of reasoning typical of a decision maker endowed with intentions 
(von Glasserfeld 1991, 1996; von Foerster 1999, 2003).

\subsection{Relation between stability and probability}

Any complex system, under given external conditions, tends to occupy the
most stable state. At the same time, the available states can be classified 
by their probabilities, so that the more stable state enjoys the higher 
probability. This relation between stability and probability is widely
acknowledged for stationary systems that can be characterized by thermodynamic
potentials, as can be inferred from textbooks on statistical physics
(Khinchin 1949; Landau and Lifshitz 1980; Yukalov and Shumovsky 1990; Sornette 2006). 
The choice of a thermodynamic potential depends on the accepted thermodynamic 
variables. For instance, if one deals with the free energy $F$, then the system 
stability requires the realization of the free energy minimum. And the  
probability of a system to possess this free energy is written as 
$p \propto e^{-\beta F}$, hence the minimization of free energy is equivalent 
to the maximization of probability. It is admissible to accept the entropy $S$ 
as a thermodynamic potential. Then the system stability requires the maximization 
of the system entropy. At the same time, the system probability reads as
$p \propto e^{S}$. Therefore, the larger entropy corresponds to the higher
probability. In Sec. 5 below, we demonstrate that the same relation between 
stability and probability holds for nonequilibrium systems as well. For the 
latter systems, the higher stability is characterized by the smaller map 
multiplier, thus, by the larger probability that is inversely proportional to
this map multiplier. In summary, for all systems, the most stable state is the most
probable one.

\subsection{Self-organization as search for the most preferable state}

A complex system can be in several macrostates corresponding to different
levels of self-organization. The macrostates can be distinguished, e.g.,  
by their order parameters (Landau and Lifshitz 1980; Yukalov and Shumovsky 1990; 
Sornette 2006) or by their order indices (Coleman and Yukalov 2000; Yukalov 2002). 
Let us denote such macrostates as $\pi_j$, enumerating them by the index 
$j = 1,2, \ldots, L$. The total set of admissible states is denoted as
\be
\label{1}
\cL = \{ \pi_j: \; j=1,2,\ldots,L \} \; .
\ee

The main assumption is that each state $\pi_j$ can be characterized by the 
related probability $p(\pi_j)$ satisfying the standard properties
\be
\label{2}
 \sum_{j=1}^L p(\pi_j) = 1 \; , \qquad 0 \leq p(\pi_j) \leq 1 \; .
\ee
For a while, we assume that such a probability measure can be defined. And
the method of constructing the corresponding probabilities will be given in 
the following section. 

Here, the index $j$, enumerating the macrostates, is taken to be discrete. 
The case of a continuous index can be treated in the same manner, just 
replacing summation by integration.   
 
The set of all states can be ordered according to relations between the
corresponding probabilities. The state $\pi_1$ is said to be preferred to
the state $\pi_2$ if and only if
\be
\label{3}
p(\pi_1) > p(\pi_2) \qquad ( \pi_1 > \pi_2 ) \; .
\ee
The states $\pi_1$ and $\pi_2$ are equivalent if and only if
\be
\label{4}
 p(\pi_1) = p(\pi_2) \qquad ( \pi_1 = \pi_2 ) \;  .
\ee
And the state $\pi_1$ is preferred or indifferent to $\pi_2$ when
\be
\label{5}
 p(\pi_1) \geq p(\pi_2) \qquad ( \pi_1 \geq \pi_2 ) \;  .
\ee

Comparing the states by using their probabilities makes it possible to define 
the least preferred and the most preferred states, which makes the set (\ref{1}) 
a complete lattice. An optimal state $\pi_*$ is the state possessing the 
largest probability:
\be
\label{6}
 p(\pi_*) \equiv \sup_j p(\pi_j) \;  .
\ee

A self-organizing complex system behaves as if it would evaluate, by means of 
fluctuations, the probabilities of available macrostates, selecting from them the 
optimal state. This is analogous to the behavior of a decision maker who chooses, 
by deliberation, among a set of given alternatives, the optimal prospect.

\subsection{Measures of system self-organization}

To define the level of self-organization, one usually considers the Shannon 
entropy
\be
\label{7}
  S \equiv - \sum_{j=1}^L p(\pi_j) \ln p(\pi_j) \; ,
\ee
which is a positive quantity characterizing missing information 
(Shannon and Weaver 1949). Respectively, the Shannon information is minus 
the Shannon entropy,
\be
\label{8}
 I_S \equiv  \sum_{j=1}^L p(\pi_j) \ln p(\pi_j) \;  .
\ee
The latter quantity, describing missing information, is negative. The larger 
the system entropy, that is, the smaller the Shannon information, the lesser 
the system self-organization.

Another measure of self-organization is the von Foerster redundancy
\be
\label{9}
 R \equiv 1 \; - \; \frac{S}{S_{max} } \; ,
\ee
in which $S_{max}$ is the maximal value of entropy (von Foerster 1995; 
Pask 1996). The Shannon entropy (\ref{7}) is maximal for the uniform 
probabilities $p(\pi_j) = 1/L$, when $S_{max} = \ln L$. The larger the 
redundancy, the higher the level of the system self-organization.    

According to von Foerster, when the system is certainly in a fixed state 
$\pi_f$, such that $p(\pi_j) = \delta_{jf}$ then it is perfectly organized, 
with $R = 1, S = 0$. On the contrary, for the uniform distribution 
$p(\pi_j) = 1/L$, the system is not organized, with $R = 0,\; S=\ln L$.

A convenient measure is the Kullback-Leibler (1951, 1959) relative 
information
\be
\label{10}
 I_{KL} \equiv  \sum_{j=1}^L p(\pi_j) 
\ln \;\frac{p(\pi_j)}{p_0(\pi_j)} \;  ,
\ee
where $p_0(\pi_j)$ is a representative of an approximate, or trial, 
probability measure, based on the available additional information on 
the system, and satisfying the standard conditions
\be
\label{11}
 \sum_{j=1}^L p_0(\pi_j) = 1 \; , \qquad 
0 \leq p_0(\pi_j) \leq 1 \;  .
\ee
The Kullback-Leibler relative information, also called negentropy, 
is non-negative defined:
\be
\label{12}
 I_{KL} \geq 0 \;  .
\ee
This follows from the Gibbs-Klein (Gibbs 1902; Klein 1931) inequality
\be
\label{13}
\sum_j  p(\pi_j) \ln \;\frac{p(\pi_j)}{p_0(\pi_j)} \; \geq \;
\sum_j \; [ p(\pi_j) - p_0(\pi_j) ]  = 1 -1 =0 \; .
\ee
The relative information is minimal when $p_0(\pi_j)$ and $p(\pi_j)$
coincide,
\be
\label{14}
I_{KL} = 0 \; , \qquad p_0(\pi_j) = p(\pi_j) \;   .
\ee
In the case of the uniform trial distribution $p_0(\pi_j)$, the 
Kullback-Leibler relative information and Shannon information are connected 
by the equality
\be
\label{15}
 I_{KL} = I_S + \ln L \; , \qquad p_0(\pi_j) = \frac{1}{L} \; .
\ee

The form of the Kullback-Leibler information is similar to the 
expected log-likelihood function employed in statistics (Edwards 1972).  
The information measures are important for constructing the information 
functional that makes it possible to define the state probabilities for
the considered complex systems.

\section{Principle of minimal information}

A pivotal role for defining the explicit form of the state probabilities is
played by the principle of minimal information implying the minimization of
an information functional. The origin of this principle is the maximization 
of entropy under given conditions 
(Gibbs 1902, 1928, 1931; Shannon and Weaver 1949; Janes 1957). The
minimization principle defines the most accurate distribution under the minimal 
available information on the considered system.

\subsection{Minimization of the information functional}

To define an information functional, one has, first, to introduce the 
representative ensemble, which is a pair $\{\mathcal{L}, p\}$, where $p$ 
implies the probability set 
$$
 p \ra \{ p(\pi_j) : \; \pi_j \in \cL ; \; j = 1,2,\ldots,L \} \; ,
$$
which is complemented by the available additional constraints making unique 
the system description (Gibbs 1928, 1931; Yukalov 1991, 2007). Such constraints 
are formulated as statistical averages, or expected values, of constraint functions:
\be
\label{16}
C_\al = \sum_j p(\pi_j) C_\al(\pi_j) \; ,
\ee
with the index $\alpha = 1,2, \ldots$ enumerating the constraints. Then, the 
information functional can be written as
$$
I[p] = \sum_j p(\pi_j) \ln \; \frac{p(\pi_j)}{p_0(\pi_j)} \; + \;
\lbd_0 \left [ \sum_j p(\pi_j) - 1 \right ] \; +
$$
\be
\label{17}
 + \; \sum_\al \lbd_\al \left [ \sum_j p(\pi_j) C_\al(\pi_j) - C_\al 
\right ] \;  ,
\ee
where $\lambda_0$ and $\lambda_\alpha$ are Lagrange multipliers guaranteeing 
the validity of constraints (\ref{16}). 

The information functional is the sum of the Kullback-Leibler information measure 
and of those constraints that have been imposed on the system. 

The minimization of the information functional assumes the variational conditions
\be
\label{18}
  \frac{\dlt I[p]}{\dlt p(\pi_j)} = 0 \; , \qquad
\frac{\dlt^2 I[p]}{\dlt p(\pi_j)^2} > 0 \; .
\ee
Introducing the global constraint
\be
\label{19}   
C(\pi_j) \equiv \sum_\al \lbd_\al C_\al(\pi_j) \; ,
\ee
this results in the state probability
\be
\label{20}
 p(\pi_j) = \frac{p_0(\pi_j)}{Z} \; \exp \{ - \beta C(\pi_j) \} \;  ,
\ee
in which the normalization quantity
$$
Z = \sum_j p_0(\pi_j) \exp\{ - \beta C(\pi_j) \} 
$$
is called partition function. The parameter $\beta$ is a Lagrange multiplier.

The meaning of the principle of minimal information is in characterizing 
the probability distribution under the minimal information encoded in the
statistical constraints. By specifying these constraints for concrete  
systems, one gets particular forms of the probability distribution.

\subsection{Minimization of the grand potential}

It may happen that the probabilities (\ref{20}) depend on some additional
set of parameters $w \ra \{w_j\}$, so that the state probability is  
\be
\label{21}
 p(\pi_j,w) = \frac{p_0(\pi_j,w)}{Z(w)} \; \exp \{ - \beta C(\pi_j,w) \} \;   ,
\ee
with the partition function 
$$
 Z(w) = \sum_j  p_0(\pi_j,w) \exp \{ - \beta  C(\pi_j,w) \} \;  .
$$
Substituting this into the information functional (\ref{17}) yields
\be
\label{22}
I[p,w] = - \ln Z(w) - C \; ,
\ee
with a fixed global constraint
$$
C \equiv \sum_\al \lbd_\al C_\al \;  .
$$
Since the latter is fixed, the minimization of the information functional
with respect to the parameter set $w$ is equivalent to the maximization of 
the partition function:
\be
\label{23}
 \min_w I[p,w] \longleftrightarrow \max_w Z(w) \;  .
\ee  
We can introduce the grand potential 
\be
\label{24}
 \Om(w) \equiv - \;\frac{1}{\beta}\; \ln Z(w) \;  .
\ee
In a thermodynamical system, $1/\beta$ plays the role of temperature $T$. 
More generally, $T$ plays the role of a parameter measuring the level 
of noise. Then, from Eq. (\ref{23}), it follows (Yukalov 2011) that the 
minimization of the information functional is equivalent to the 
minimization of the grand potential:
\be
\label{25}
 \min_w I[p,w] \longleftrightarrow \min_w \Om(w) \;  .
\ee

The above formalism applies beyond the description of thermodynamical 
systems at or close to equilibrium, also to quasi-equilibrium systems, 
when the notion of temperature is replaced by its more generalized version 
$1/\beta$ giving a measure of the typical strength of the fluctuations 
of the system variables. As important practical applications, it is 
possible to enumerate a number of heterogenous condensed-matter systems 
displaying mesoscopic heterophase fluctuations (Yukalov 1981, 1991, 2003a; 
Shumovsky and Yukalov 1982). Then the parametric set $\{w_j\}$, normalized 
in the usual way,
$$
 \sum_j w_j = 1 \; , \qquad 0 \leq w_j \leq 1 \;  ,
$$
characterizes the statistical weights $w_j$ of qualitatively different 
mesoscopic fluctuations.

\section{Quasi-stationary self-organizing systems}

When the system parameters vary much slower than typical dynamical 
motions in the system, the latter can be treated as quasi-stationary. 
The principle of minimal information is often applied for describing 
self-organization in such quasi-stationary systems. Below we give a 
brief reminder of the known examples of various statistical systems,
stressing the main idea that the systems of quite different nature can 
be described in a general probabilistic way.

\subsection{Probability of thermodynamic states}

Statistical systems, to which thermodynamics is applicable, are characterized 
by thermodynamic potentials, such as the free energy (Landau and Lifshitz 1980).
Suppose that the system can acquire several thermodynamic states $\pi_j$, 
corresponding to different thermodynamic phases specified by different order 
parameters and the related symmetry (Landau and Lifshitz 1980; 
Yukalov and Shumovsky 1990; Coleman and Yukalov 2000; Sornette 2006). For each 
such a state, one can define a thermodynamic potential, for concreteness, the 
free energy $F(\pi_j)$. Then, as a constraint (\ref{16}), it is natural to take 
the expected  value
\be
\label{26}
 \overline F = \sum_j p(\pi_j) F(\pi_j) \;  .
\ee
Assuming the uniform trial distribution $p_0(\pi_j) = 1/L$, the principle of
minimal information gives  
\be
\label{27}
 p(\pi_j) = \frac{1}{Z} \; \exp\{ - \bt F(\pi_j) \} \;  ,
\ee
with the partition function
$$
 Z = \sum_j \exp \{ - \bt F(\pi_j) \} \; .
$$
Here $\beta$ is a Lagrange multiplier corresponding to the inverse temperature
$T = 1/\beta$.

Distribution $p(\pi_j)$ describes the probability that the thermodynamic system 
is in the state $\pi_j$. The sharpness of the distribution depends on the 
system size.

\subsection{Infinite statistical systems}

The typical situation for the so-called bulk statistical systems,
such as condensed matter or gases, is to consider their large sizes by 
taking the thermodynamic limit, when the number of particles $N$ composing  
the system is assumed to tend to infinity. In that case, the free energy, being 
an extensive quantity, tends to infinity as $F(\pi_j) \propto N \ra \infty$. 
Therefore the probability (\ref{27}) becomes sharply centered at the optimal 
state $\pi_*$,
\begin{eqnarray}
\label{28}
p(\pi_j) = \left \{ \begin{array}{ll}
1 , & ~~ \pi_j = \pi_* \\
0 , & ~~ \pi_j \neq \pi_* 
\end{array} \; , \right.
\end{eqnarray}
whose order parameter provides the minimal free energy of the system,
\be
\label{29}
 F(\pi_*) = \min_j F(\pi_j) \; .
\ee

Let some of the characteristic system parameters, either external or internal, 
be varying. For instance, this can be temperature. For any given governing 
parameter, such as temperature, the system always chooses the optimal state, 
as described above. As an illustration, let us vary the temperature and let there 
exist two different states, corresponding to different thermodynamic phases.
Then, there can exist a critical temperature at which the following transition 
occurs,
$$
p(\pi_1) = 1 \qquad ( T < T_c) \; ,
$$
\be
\label{30}
p(\pi_2) = 1 \qquad ( T > T_c) \;   ,
\ee
which is called phase transition.  

Since varying the governing parameters does not usually directly impose 
neither the type of the order parameter nor the related symmetry, but the 
system itself acquires the structure and symmetry of the optimal state, 
this process is termed self-organization. As far as the system takes the 
optimal state with probability one, the process of self-organization for 
an infinite system is of the deterministic type. Counter examples are 
provided by spin glasses and other so-called ill-condensed systems, which 
exhibit the coexistence of exponentially many probabilistically almost 
equivalent states even in the thermodynamic limit (M\'ezard et al. 1986).

\subsection{Finite statistical systems}

There exists a large class of systems that contain many particles, in that 
sense being statistical, but at the same time, with the number of 
particles being finite, such that finite-size effects become important.
Examples are trapped atoms, quantum dots, atomic nuclei, metallic grains,
and spin assemblies, as well as biological molecular structures 
(Birman et al. 2013).

When the system is finite, with a finite number $N$ of particles, then 
several macroscopic states can be realized, having nontrivial probabilities 
(\ref{27}). In such a case, the phase transition between two different phases 
occurs with the characteristics that the state probabilities are not exactly 
one or zero, as for infinite statistical systems. The transition now is 
characterized by the reversion of the inequality
\be
\label{31}
 p(\pi_1) > p(\pi_2)  \qquad ( T < T_c)  
\ee
to the inequality
\be
\label{32}
  p(\pi_1) < p(\pi_2)  \qquad ( T > T_c) \;  .
\ee
The transition can be discontinuous or continuous. In any case, this corresponds 
to a probabilistic self-organization associated with phase transitions 
(Bouchaud and Georges 1990; Jona-Lasinio 2001).

As physical illustrations of finite systems with coexisting phases, we can 
mention metallic grains that can be either in superconducting or normal states 
(von Delft 2001), atomic nuclei that can take different shapes (Gaudefroy et al. 2009),
and nanosize spin clusters that can be either in magnetic or non-magnetic states 
(Bansmann et al. 2005).

\subsection{Financial and social systems}

The methods of statistical physics and thermodynamics have been widely 
used for economic, financial, and social systems, as can be inferred from
the reviews (Baumg\"{a}rnter 2004; Smith and Foley 2008; Castellano et al. 2009;  
Yakovenko and Rosser 2009). A statistical description, characterized by 
the probabilities of type (\ref{20}), has been employed for various 
financial and social systems, with different quantities playing the role of 
constraints $C_\alpha(\pi_j)$. Foley (1994, 1996) applied such a probability 
distribution for financial markets, treating the constraints as market 
transactions of different agents and using the term {\it market temperature}. 
The transaction values can be measured in money units (Yakovenko and Rosser 2009; 
Kusmartsev 2011). Market crashes can be associated with phase transitions 
(Sornette 2003; McCauley 2003; Marsili 2009). As {\it market energy}, or 
{\it disagreement function}, one often uses Hamiltonians of spin systems 
(Chowdhury and Stauffer 1999; Zhou and Sornette 2007; Stauffer 2008;
Harras and Sornette, 2011). In applications to social systems, one uses a 
constraint that is equivalent to the system energy and is called 
{\it system frustration}, or {\it conflict} (Galam and Moscovici 1991; Galam 1996; 
Florian and Galam 2000; Gallo et al. 2009).

Markets or social groups are, certainly, finite systems, hence, they can 
be characterized by the distributions of type (\ref{27}). The free energy
for financial systems can be defined (Smith and Foley 2008) as ``an intrinsic
money-metric welfare measure of the allocation of an economy in contact with
a reservoir". 

Extending the definition of free energy to social systems, one defines it in 
the following way. The system energy $E(\pi_j)$ of a state $\pi_j$ can be 
termed the {\it state cost}. The society temperature $T$ has the meaning of the 
intensity of noise produced by the surrounding playing the role of a thermostat. 
The noise energy, or the cost of noise for a system in a state $\pi_j$, is given 
by the quantity $TS(\pi_j)$, where $S(\pi_j)$ is the entropy of the state $\pi_j$. 
Then, the free energy, or free cost, is the intrinsic cost of a state, that is, 
the cost of the state without the cost of noise: 
\be
\label{33}
F(\pi_j) = E(\pi_j) - TS(\pi_j) \;   .
\ee

Being a finite system, a society cannot be in a single pure state, but always
possesses finite probabilities of being in different states. Phase transitions
occur from one dominant state to another, as is described in Sec. 4.3. 
Continuous transitions correspond to fast evolutions, while discontinuous 
transitions are associated with revolutions or abrupt regime shifts.

\subsection{Biological and ecological systems}

For biological and ecological systems, $\pi_j$ can correspond to a type of 
species characterized by fitness $w(\pi_j)$. The intensity of external noise
is described by {\it selection temperature} $T$. As constraint (\ref{16}), 
one defines the average fitness
\be
\label{34}
 W = \sum_j p(\pi_j) w(\pi_j) \;  .
\ee
Then, from the principle of minimal information of Sec. 3, one finds the 
distribution called the {\it relative reproduction rate}
\be
\label{35}
 p(\pi_j) = \frac{1}{Z} \; \exp \{ \bt w(\pi_j) \} \;  .
\ee
This exponential form of the reproduction rate is often used in the 
biological literature (Manly 1976; Crozier and Pamilo 1979; Russell 1996; 
Arias et al. 2001; Cowperthwaite et al. 2005; Martin and Lenormand 2008; 
Saakian et al. 2010). Expression (\ref{35}) shows that, among a variety of 
different species, the one with highest fitness enjoys the higher 
reproduction rate. 

\vskip 2mm

The goal of this section has been the demonstration of the main idea that 
rather different systems can be described in a general probabilistic way
enjoying the same mathematical characterization.

\section{Self-organization in dynamical systems}

In dynamical systems, self-organization is usually accompanied by the 
appearance of spatial structures or patterns (Glansdorff and Prigogine 1971;
Haken 1983, 2005; Nicolis 1986) or it is connected with critical
transitions (Kuehn 2011) when the system behavior changes qualitatively.
The process of self-organization in dynamical systems can also be formulated
in a probabilistic framework (Yukalov 2001a, 2001b, 2003b). 

The main message of the present section is twofold. First, we demonstrate that 
nonequilibrium systems, similarly to equilibrium ones, can be characterized by
probabilities derived from the principle of minimal information, that is, from 
conditional entropy maximization. Second, we prove that the notion of stability
is directly connected to that of probability. The more stable state is described 
by the smaller map multiplier and by the larger probability.

\subsection{Probabilistic pattern selection}

Suppose a dynamical system can acquire several different spatial structures,
with the type of the $j$-th structure being denoted by $\pi_j$. To make the 
description of the probabilistic approach transparent, let us consider a 
one-dimensional dynamical system, whose evolution is given by the equation 
\be
\label{36}
 \frac{d}{dt} \; y(\pi_j,t) = v(\pi_j,t) \;  ,
\ee 
where $y(\pi_j,t)$ represents an observable quantity. A generalization to 
dynamical systems of any dimensionality is straightforward 
(Yukalov 2001a, 2001b, 2003b).

Self-organization of a dynamical system can be interpreted as the system 
search for stability. The latter is characterized by the {\it map multipliers}
\be
\label{37}
 \mu(\pi_j,t) \equiv \frac{\dlt y(\pi_j,t)}{\dlt y(\pi_j,0)} \; .
\ee
If $|\mu(\pi_j,t)| < 1$, the structure $\pi_j$ is locally stable at time $t$.
When $|\mu(\pi_j,t)| = 1$, the structure is locally neutral, and when
$|\mu(\pi_j,t)| > 1$, the structure is locally unstable. The map multiplier
can be expressed through the Jacobian 
\be
\label{38}
J(\pi_j,t) \equiv \frac{\dlt v(\pi_j,t)}{\dlt y(\pi_j,t)} \;
\ee
in the form
\be
\label{39}
 \mu(\pi_j,t) = \exp \left \{ \int_0^t J(\pi_j,t') \; dt' 
\right \} \;  .
\ee

It is convenient to introduce the {\it expansion exponents}
\be
\label{40}
 X(\pi_j,t) \equiv \ln | \mu(\pi_j,t) | \;  .
\ee
These exponents show how quickly a deviation from an initial condition 
varies in time, either converging to or diverging from this initial 
condition according to the relation
$$
| \dlt y(\pi_j,t) | = | \dlt y(\pi_j,0) | \exp \{ X(\pi_j,t) \} \;   .
$$
The expansion exponent is connected with the Jacobian by the equation
\be
\label{41}
 X(\pi_j,t) = {\rm Re} \int_0^t J(\pi_j,t') \; dt' \;  .
\ee
  
Our aim is to find an expression for the structure probability $p(\pi_j,t)$
that should satisfy the standard normalization condition
\be
\label{42}
 \sum_j p(\pi_j,t) = 1 \; , \qquad 0 \leq p(\pi_j,t) \leq 1 \;  ,
\ee
at each moment of time. Following the general prescription of Sec. 3, we 
define the information functional 
\be
\label{43}
 I[p] = \sum_j p(\pi_j,t) \ln \; \frac{p(\pi_j,t)}{p_0(\pi_j,t)} \; +
\; \lbd_0 \left [ \sum_j p(\pi_j,t) - 1 \right ] \;  .
\ee
By the assumption that the system searches for the most stable structure,
the trial distribution $p_0(\pi_j,t)$ can be taken to be inversely 
proportional to the modulus of the map multiplier $|\mu(\pi_j,t)|$. Then,
from the principle of minimal information, we get the structure probability
\be
\label{44}
p(\pi_j,t) = \frac{1}{Z(t)|\mu(\pi_j,t)|} \;   ,
\ee
with the partition function
$$
Z(t) = \sum_j \frac{1}{ |\mu(\pi_j,t)|} \;  .
$$
In view of relation (\ref{40}), the structure probability can be expressed
through the expansion exponent as
\be
\label{45}
 p(\pi_j,t) = \frac{\exp\{-X(\pi_j,t)\}}{Z(t)} \;  ,
\ee
where the partition function is
$$
Z(t) = \sum_j \exp\{ -X(\pi_j,t) \}\;   .
$$

Thus, the dynamical system, in general, can exhibit different structures,
with the corresponding probabilities (\ref{45}). The system tries to 
self-organize acquiring the most stable structure. At the same time, other
less stable structures are also admissible, though with lower probabilities. 
Between two structures at the given moment of time, the structure that is 
more probable is the one which is more stable and whose expansion exponent 
is smaller. This can be called the {\it principle of minimal expansion}. 
This approach can be employed for any dynamical system. Time series, met in 
various empirical data, can be represented as trajectories of dynamical 
systems. Therefore, the approach of pattern selection can be applied to 
different time series as well, e.g., to time series that are commonly found 
for financial markets (Yukalov 2001c).   

The importance of the present section is the direct demonstration of the intimate 
relation between stability and probability for nonequilibrium systems. 
{\it The most stable state is the most probable}.

\subsection{Turbulent photon filamentation}

In order to show that the results of the previous section provide a practical tool
for treating concrete nonequilibrium systems, let us consider the effect 
of turbulent photon filamentation. This is the phenomenon in which an assembly 
of resonant atoms inside a cylindrical sample spontaneously separates into many 
thin radiating filaments that are randomly distributed in the sample cross-section 
(Encinaz-Sanz et al. 2000, Leyva and Guerra 2002). Similar random structures 
also arise in passive nonlinear media, such as Kerr media, and in active nonlinear 
media, such as photorefractive crystals, pumped by a uniform laser beam 
(Arecchi et al. 1999).    

The microscopic description for a system made of two-level resonant atoms starts 
with the Hamiltonian
\be
\label{46}
 \hat H = \hat H_a + \hat H_f + \hat H_{af} \; ,
\ee
consisting of the atomic Hamiltonian $\hat{H}_a$, field Hamiltonian 
$\hat{H}_f$, and the Hamiltonian of atom-field interactions, $\hat{H}_{af}$.
The atomic Hamiltonian is 
\be
\label{47}
\hat H_a = \sum_{i=1}^N \om_0 \left ( \frac{1}{2} + S_i^z \right ) \; ,
\ee
where $N$ is the number of atoms, $\omega_0$ is the atomic transition frequency, 
and $S_i^z$ is the $z$ - component of the pseudospin operator of the $i$-th atom. 
The field Hamiltonian is
\be
\label{48}
 \hat H_f = \frac{1}{8\pi} \int \left ( \bE^2 + \bH^2 \right ) \; d\br \;  ,
\ee
with electric field $\bE$, magnetic field $\bH ={\bf \nabla}\times{\bf A}$, 
and vector potential ${\bf A}$. The atom-field interaction is given by the 
Hamiltonian
\be
\label{49}
 \hat H_{af} = - \; \frac{1}{c} \sum_{i=1}^N {\bf J}_i \cdot {\bf A}_i \;  ,
\ee
in which the short-hand notation ${\bf A}_i \equiv {\bf A} (\br_i, t)$
is used and the transition current has the form
\be
\label{50}
 {\bf J}_i = i \om_0 \left ( {\bf d} S_i^+ - {\bf d}^* S_i^- \right ) \;  ,
\ee
where ${\bf d}$ is the transition dipole and $S_i^{\pm}$ are the ladder 
operators.

The dynamical system is composed of the evolution equations for the average
vector potential $\langle {\bf A}\rgl$ and the pseudospin averages describing
dipole transitions,
\be
\label{51}
 u_i(t) = 2 \lgl S_i^-(t) \rgl \; ,
\ee
coherence intensity
\be
\label{52}
  w_i(t) \equiv \frac{2}{N} \sum_{j(\neq i)} \; \lgl S_i^+(t) S_j^-(t)
+  S_j^+(t) S_i^-(t) \rgl \;  ,
\ee
and the population difference
\be
\label{53}
 s_i(t) \equiv 2 \lgl S_i^z(t) \rgl \;  .
\ee

The arising filaments can possess different radii $r_j$ that correspond 
to different structures. Employing the method of Sec. 5.1, it is possible 
to find (Yukalov  2000, 2001a, 2001b) that the probability of a filamentary 
structure with the radius $r_j$ has the form
$$
p(r_j,t) = \frac{1}{Z} \; \exp\left\{ - {\rm Re} 
\int_0^t {\rm Tr} \hat J(r_j,t') \; dt' \right \} \;   ,
$$
with
$$
{\rm Tr} \hat J(r_j,t) = -\gm_1 -\gm_3 - 2\gm_2 [\; 1 - g(r_j) s \; ] \; ,
$$
where $\gamma_1, \gamma_2, \gamma_3$ are the longitudinal, transverse, and 
dynamical attenuations, respectively, $g(r_j)$ is the effective atomic
interaction in the related structure with the filament radius $r_j$, and
$s$ is the average population difference (\ref{53}) in a filament of that structure.       

The maxima of the above probability define the filament radii corresponding 
to the zeroes of the integral sine:
\be
\label{54}
Si \left ( \frac{4\pi\sqrt{e}}{\lbd L} \; r_j^2 \right ) = 0 \;   ,
\ee
where $\lambda$ is the radiation wavelength and $L$ is the length of the 
sample. The optimal radius is given by the absolute maximum of the structure
probability, yielding 
\be
\label{55}
r_* = 0.3 \sqrt{\lbd L} \;   .
\ee

The majority of the arising filaments have this radius (\ref{55}), although
the filaments with other radii, satisfying Eq. (\ref{54}), are also present. 
These theoretical results have been found to be in very good agreement with 
experiments (Encinaz-Sanz et al. 2000, Leyva and Guerra 2002).

\section{Decision making as self-organization}

In the examples treated above, we have shown that practically any system,
whether natural, social, financial, biological or ecological, can be 
characterized by a probability measure prescribing a weight to each admissible
system state or structure. The larger the state probability, the more stable 
the system, and the more often the state is realized. The process of 
self-organization works as if the system would be searching for the most stable 
state corresponding to the largest probability. In the same way, the process
of decision making can be interpreted as self-organization in the decision 
maker nervous system, when the decision maker is searching for the most 
preferable prospect, which thus can be seen as the most stable, characterized 
by the largest probability.

\subsection{Classical utility theory}

Classical decision theory is based on the notion of expected utility
(von Neumann and Morgenstern 1953; Savage 1954). We briefly recall the basic
definitions that will be used in what follows. 

The consequences of actions are measured by outcomes, or payoffs, composing 
a set    
\be
\label{56}
 \mathbb{X} \equiv \{ x_n \in \mathbb{R} : \; 
n = 1,2,\ldots,N_{out} \} \;  .
\ee
The payoffs can be weighted in different ways, by means of different 
probability measures over the set (\ref{56}), enumerated with the index 
$j = 1,2,\ldots,L$, with the probabilities $\{p_j(x_n)\}$ satisfying the 
standard normalization condition
\be
\label{57}
\sum_{n=1}^{N_{out} } \; p_j(x_n) = 1 \; , \qquad
0 \leq p_j(x_n) \leq 1 \; .
\ee

A lottery, is the set of payoffs and their weights,
\be
\label{58}
\pi_j = \{ x_n , p_j(x_n) : \; n= 1,2, \ldots, N_{out} \} \;   .
\ee
One defines the lottery mean
\be
\label{59}
\overline x(\pi_j) \equiv 
\frac{1}{N_{out}} \; \sum_{n=1}^{N_{out}} x_n p_j(x_n) 
\ee
and the lottery variance
\be
\label{60}
 {\rm var}(\pi_j) \equiv 
\frac{1}{N_{out}} \; \sum_{n=1}^{N_{out}} \left( x_n^2 p_j(x_n) \; -
\; \left [ \overline x(\pi_j) \right]^2 \right) \;  .
\ee
One calls the lottery uncertain when its variance is not zero, and it is
certain if the variance vanishes. 

On the set of payoffs, one defines a utility function 
$u(x): \mathbb{X} \ra \mathbb{R}$, which is non-decreasing and concave
(Bernoulli 1738). The cardinal expected utility reads as
\be
\label{61}
 U(\pi_j) = \sum_{n=1}^{N_{out}} u(x_n) p_j(x_n) \;  .
\ee

The expected utility serves as a characteristic of the lottery usefulness.
 
One says that a lottery $\pi_1$ is more useful than $\pi_2$, if and only if
\be
\label{62}
 U(\pi_1 ) > U(\pi_2) \;  .
\ee
Two lotteries are equally useful, when
\be
\label{63}
  U(\pi_1 ) = U(\pi_2) \;  .
\ee
And a lottery $\pi_1$ is not less useful than $\pi_2$ if
\be
\label{64}
  U(\pi_1 ) \geq U(\pi_2) \;  .
\ee

The action of choosing a lottery under uncertainty is termed a {\it prospect}.  
The prospects are analogous to the system states considered in Sec. 2. The 
set (\ref{1}) of all admissible prospects, which are ordered according to 
relations (\ref{62}) to (\ref{63}), is termed a {\it lattice}. Among all 
prospects, there exists the least useful one, $\pi_{min}$, whose expected 
utility is the smallest:
\be
\label{65}
  U(\pi_{min} ) =\min_j U(\pi_j) \;  .
\ee
And there is the most useful prospect $\pi_{max}$, with the largest expected 
utility:
\be
\label{66}
  U(\pi_{max} ) =\max_j U(\pi_j) \; .
\ee
Because of this, the prospect set corresponding to (\ref{58}) forms a 
{\it complete lattice}.  

In this formulation, classical decision theory is deterministic since
a decision maker is supposed to necessarily prefer the most useful prospect.

\subsection{Probabilistic utility theory}

The classical normative utility theories as well as different descriptive 
behavioral utility theories, such as prospect theory 
(Tversky and Kahneman 1973; 1980; 1983), are all deterministic, requiring, 
with certainty to prefer the prospect characterized by the largest functional 
quantifying the considered prospects. However, as we show below, decision 
theory can be reformulated in probabilistic language.    

Our aim is to describe the process of decision making as an intrinsically 
probabilistic procedure. The first step consists in evaluating consciously 
and/or subconsciously the probabilities of choosing different actions from the 
point of view of their usefulness and/or appeal to the choosing agent. We 
transform the above classical deterministic approach to the general 
probabilistic formulation by assuming that the prospects $\pi_j$ are not fixed, 
but represent random variables, so that the prospect lattice is a field of 
random events (Luce 1958). Respectively, the expected utility (\ref{61}) is 
also a random quantity that can be characterized by a distribution of prospects 
$f(\pi_j)$, with the usual normalization condition
\be
\label{67}
 \sum_{j=1}^L f(\pi_j) = 1 \; , \qquad 0 \leq f(\pi_j) \leq 1 \; .
\ee
The weight $f(\pi_j)$ can be called the {\it utility factor}, since it
describes the usefulness of the prospect $\pi_j$. According to this meaning,
the usefulness of a prospect with zero utility has to be zero, which imposes
the limiting condition 
\be
\label{68}
 f(\pi_j) \ra 0 \; , \qquad  U(\pi_j) \ra 0 \;  .
\ee
Being a random quantity, the utility $U(\pi_j)$ is assumed to be normalized as
\be
\label{69}
 \sum_{j=1}^L f(\pi_j) U(\pi_j) = U \; .
\ee
This practical condition guarantees that the involved lotteries are well 
defined, having finite expected utilities.
 
To find the distribution $f(\pi_j)$, we resort to the principle of minimal 
information of Sec. 3, introducing the information functional
$$
I[f] = \sum_j f(\pi_j) \ln \; \frac{f(\pi_j)}{f_0(\pi_j)} \; + \;
\lbd \left [ \sum_j f(\pi_j) - 1 \right ] \; -
$$
\be
\label{70}
 - \; \bt \left [ \sum_j f(\pi_j) U(\pi_j) - U \right ] \;  ,
\ee
with the Lagrange multipliers $\lambda$ and $\beta$ taking into account 
conditions (\ref{67}) and (\ref{69}). To satisfy condition (\ref{68}), the 
trial distribution $f_0(\pi_j)$ can be defined as the likelihood ratio 
proportional to $U(\pi_j) / U(\pi_{max})$. Then the minimization of the 
information functional leads to the utility factor
\be
\label{71}
 f(\pi_j) = \frac{ U(\pi_j)}{Z} \; \exp\{ \bt U(\pi_j) \} \; ,
\ee
with the partition function
$$
Z =  \sum_j U(\pi_j) \exp\{ \bt U(\pi_j) \} \; .
$$
Note that the utility factor specified by (\ref{71}) satisfies the 
limiting condition (\ref{68}). 

A probabilistic representation of decisions is not new, since it is at the core 
of classical choice theory (Anderson et al. 1992). Classical choice theory 
assumes that the probability to choose between different alternatives can be written 
similarly to expression (\ref{71}) but without the $U(\pi_j)$ prefactor, which is 
called the logit rule (McFadden 1974). While similar (in particular with the use 
of entropy arguments), our formulation is essentially different from the logit rule
and this difference results from the specification (\ref{68}).

The Lagrange multiplier $\beta$ plays the role of a parameter capturing
the level of confidence or belief in selecting the prospects, hence, $\beta$
can be called the {\it belief parameter} or {\it confidence parameter}. 
Requiring that the utility factor be an increasing function of utility 
makes the belief parameter non-negative, $\beta \geq 0$. The limiting values 
of this parameter characterize decision making in the situations of 
underconfidence or overconfidence (Griffin and Tversky 1992). In the 
particular case of no confidence, we have
\be
\label{72}
 f(\pi_j) = \frac{ U(\pi_j)}{\sum_j U(\pi_j) } \qquad ( \bt = 0 ) \;  .
\ee
In the opposite case of extreme confidence, we get
\begin{eqnarray}
\label{73}
f(\pi_j) = \left \{ \begin{array}{ll}
1 , & ~~ \pi_j = \pi_{max} \\
0 , & ~~ \pi_j \neq \pi_{max} 
\end{array} \qquad (\bt \ra \infty) \; . \right.
\end{eqnarray}
The latter situation recovers the deterministic formulation of utility
theory of Sec. 6.1.   

The ordering of prospects by their usefulness can be done by means of the 
utility factors. A prospect $\pi_1$ is deemed more useful than $\pi_2$, 
if and only if
\be
\label{74}
 f(\pi_1) > f(\pi_2) \;  .
\ee
Two prospects are equally useful when
\be
\label{75}
 f(\pi_1) = f(\pi_2) \;  .
\ee
And a prospect $\pi_1$ is not less useful than $\pi_2$ if
\be
\label{76}
f(\pi_1) \geq f(\pi_2) \;   .
\ee
This ordering is in agreement with that of Sec. 6.1, based on the 
comparison of expected utilities.  

The application of this approach to time-dependent processes is 
straightforward. This simply requires including time dependence into the 
definition of expected utility by incorporating in it a temporal discount
rate (Samuelson 1937; Loewenstein and Thaler 1989; Frederick et al. 2002; 
Rambaud and Torrecillas 2005; Berns et al. 2007). It is also possible to 
vary the definition of utility by taking into account the effects of 
aspiration and adaptation (Selten 1998; Napel 2003).      

This probabilistic formulation of utility theory puts it in the same frame
as the description of any self-organizing system presented in previous 
sections. In this framework, the process of decision making can be understood 
as the search for the most preferable prospect that enjoys the largest 
probability. This can be interpreted as describing the process of 
self-organization in the nervous system of the decision maker.

\subsection{Behavioral and quantum decision making}

Decision theory, based on utility theory, even in the probabilistic variant, 
characterizes the objective features of the involved prospects, leaving aside
all subjective effects connected with decision makers. Classical utility 
theory assumes that decision makers are rational and able to precisely estimate 
the corresponding utilities of the considered prospects. This, however, is 
a simplification of the real life, where decision makers are always subject 
to subconscious feelings, emotions, various biases, prejudices, incentives, 
anxiety, and intuitive heuristics (Tversky and Kahneman 1983; Yates and Carlson 1986;
Maturana 1988; Shafir et al. 1990; Dixit and Besley 1997; Epstein 2008; 
West and Grigolini 2010). Variants of decision theory that try to take 
account of these subjective effects, typical of the behavior of real 
decision makers, are studied in behavioral decision making 
(Machina 2008, Simon 1959). 

Sometimes, one says that realistic decision making contains generic 
indeterminism (Nichols 2011). Remembering that similar indeterminism
is typical of quantum theory, this hints on the possibility of 
characterizing behavioral decision making by means of quantum probability 
(Lehrer and Shmaya 2006). Actually, Bohr (1933, 1958) was the first to
advocate the use of quantum theory for describing psychological processes.
There have been a number of publications discussing the necessity of 
invoking quantum probabilities for behavioral decision making 
(Lehrer and Shmaya 2006; Danilov and Lambert-Mogiliansky 2008, 2010; 
Lambert-Mogiliansky et al. 2009; Photos and Busemeyer 2009). A full 
quantitative theory introducing quantum probabilities for behavioral 
decision making has been developed by the authors 
(Yukalov and Sornette 2008, 2009a, 2009b, 2010, 2011). The approach is 
based on the mathematical theory of quantum measurements, which, as has 
been noticed by von Neumann (1955), can be interpreted as a kind of 
decision theory. 

In the present subsection, we wish to emphasize that the probabilistic 
way of constructing decision theory can be extended to behavioral decision 
making taking into account such subjective features as emotions and biases.
We shall not go into details of this approach involving quantum techniques,
which can be found in our previous papers 
(Yukalov and Sornette 2008, 2009a, 2009b, 2010, 2011). But we shall only
formulate the results. 
  
By construction, quantum theory is probabilistic. The scheme of calculating
the prospect probabilities follows the rules of defining the observable 
quantities in the quantum theory of measurement. The space of mind of a decision 
maker is described by a Hilbert space on which prospect operators $\hat P(\pi_j)$ 
are defined. These operators play the role of the operators of observables, 
whose averaging yields the observable quantities corresponding to the prospect 
probabilities:
\be
\label{79}
 p(\pi_j) \equiv {\rm Tr} \hat\rho \hat P(\pi_j) \;  ,
\ee
where $\hat{\rho}$ is a statistical trace-one operator characterizing the 
decision maker, and the trace is taken over the decision-maker space of mind. 
The prospect probabilities are normalized as in condition (\ref{2}).   

It is straightforward to show that the prospect probability (\ref{79}) is 
the sum of two terms:
\be
\label{80}
 p(\pi_j) = f(\pi_j) + q(\pi_j) \;  .
\ee
The first term corresponds to the utility factor characterizing objective
features, while the second term is due to quantum effects of coherence and 
interference, which corresponds to subjective features.   

As is known (Zurek 2003), classical theory is a particular case of quantum 
theory, corresponding to the situation when coherence effects disappear, 
which is called {\it decoherence}. In the present case, in the same way as 
for any observable in quantum theory, decoherence implies the disappearance 
of the quantum coherence term $q(\pi_j)$. Then, the remaining term $f(\pi_j)$
has to correspond to the classical utility factor described in the previous
subsection. In this way, classical decision theory is obtained as a limiting 
case of the quantum decision theory, when decoherence occurs.  

The quantum coherence, or interference, term is what distinguishes quantum
prospect probabilities from their classical counterparts. From the point of 
view of quantum theory, the arising coherence term can be ascribed to quantum
indeterminacy, being contextual. Interpreted in the contexts of behavioral 
decision theory, this term can be called {\it attraction factor}, 
characterizing subjective attitudes of a decision maker to the considered 
prospects.

Thus, the quantum prospect probability (\ref{80}) is of dual nature, 
containing the objective utility factor $f(\pi_j)$, defined in terms of the 
prospect utility, and the attraction factor $q(\pi_j)$, characterizing the 
subjective attractiveness of the prospect for the decision maker.

Despite the fact that the attraction factor embodies subjective and unconscious
components of the decision making process, it enjoys several quantitative 
properties making it possible to give quantitative predictions for the prospect 
probabilities. 

First of all, the attraction factor lies in the interval
\be
\label{83}
-1 \leq q(\pi_j) \leq 1 \; .
\ee
The normalization conditions lead to the {\it alternation property}
\be
\label{84}
 \sum_{j=1}^L q(\pi_j) = 0 \; .
\ee
And the following average estimate holds:
\be
\label{85}
 \frac{1}{L} \sum_{j=1}^L | q(\pi_j) | = \frac{1}{4} \; .
\ee

These properties allow one to make {\it quantitative} predictions for 
aggregate groups of decision makers, which are found to be in excellent
agreement with empirical data, as has been shown in our previous 
publications (Yukalov and Sornette 2009a, 2009b, 2010, 2011), where
it has been demonstrated that all paradoxes of decision making arising 
from the perspective of classical utility theory find straightforward 
resolution in the behavioral quantum approach. 

The basic idea of the present subsection is to emphasize two important
facts: 

(i) First of all, subjective behavioral phenomena in decision making cannot 
be described by minimizing an information functional, whose minimization
can provide only the objective part of the total quantum probability. But 
taking into account subjective effects requires to resort to more 
elaborate techniques of quantum theory. The same, actually, concerns 
complex quantum systems, whose description also requires the use of such
techniques, but cannot be fully described by deriving quantum probabilities
from an entropy maximization or information minimization. Thus, these 
methods are general for defining the probabilities of classical systems,
but are not sufficient for characterizing quantum systems.  

(ii) Nevertheless, even quantum systems in nature, as well as subjective effects 
in decision making, allow for a unified general procedure of calculating both
the state probabilities of quantum systems as well as the prospect probabilities 
for behavioral decision makers. In that way, we see again that there is no 
principal difference between self-organization of complex systems, including 
quantum, and the process of human decision making, even subject to behavioral 
biases and emotional feelings. Behavioral effects of decision makers can be 
interpreted as analogous to quantum fluctuations in natural systems.             

According to the behavioral interpretation of quantum decision theory 
(Yukalov and Sornette 2008, 2009a, 2009b, 2010, 2011), the process of decision 
making goes through the following steps. One fixes a set of prospects, then 
evaluates their utility and attractiveness, resulting in the evaluation of 
the prospect probabilities, and from their comparison, one defines the optimal 
prospect. This scheme can be formalized by the sequence      
\be
\label{90}
 \{ \pi_j \} \ra \{ p(\pi_j) \} \ra p(\pi_*) \;  .
\ee

But the same sequence is typical of self-organization of any system. It is 
just a matter of terminology, whether one talks of decision prospects or
system states. Decision making and self-organization are the same processes,
sometimes occurring in different systems and often times happening
in the same system as self-organization of the nervous system during the 
decision process of a human being.

\section{Summary}

Self-organization in different systems is described as a process of 
evaluation of the state probabilities in the search for the most stable 
state, hence for the state with the largest probability. Natural systems evaluate 
the admissible states by means of fluctuations. The explicit expression for 
the probability distribution follows from the principle of minimal information, 
implying the minimization of an information functional. This principle 
provides the best probability distribution, under the minimal given 
information on the system. This general scheme is applicable to systems of 
any nature, whether statistical, financial, economic, social, or biological. 
The probabilistic approach is valid for quasi-equilibrium as well as 
nonequilibrium systems. 

Decision making, formulated in a probabilistic representation, is also a 
process of evaluating the prospect probabilities, in the search for an 
optimal prospect, having the largest probability. Decision makers evaluate 
the admissible prospects by deliberations. In classical decision theory, 
the prospects are classified as more or less useful according to their 
expected utilities. In the behavioral application of quantum decision making, 
the prospects are evaluated by their utility as well as by their 
attractiveness. 

In all cases, the procedure of self-organization is analogous to that of 
decision making, both being characterized by the same mathematical scheme. 
It is only the language that is slightly different. But there is a direct 
translation of one language onto another, which is exemplified in the following 
dictionary.

\begin{center}
\begin{tabular}{ll} \\
Complex system       & \hspace{1cm}  Decision maker \\
System states        & \hspace{1cm}  Decision prospects \\
System fluctuations  & \hspace{1cm}  Decision-maker deliberations \\
State probability    & \hspace{1cm}  Prospect probability \\
System stability     & \hspace{1cm}  Prospect preferability  \\
Most stable state    & \hspace{1cm}  Most preferable prospect \\
Quantum fluctuations & \hspace{1cm}  Behavioral biases \\
Self-organization    & \hspace{1cm}  Decision making  \\
\end{tabular}
\end{center}

It is possible to state that self-organization and decision making are 
equivalent processes. This conclusion is not merely important from the general 
descriptive point of view, but it has far-reaching practical consequences. 
For instance, these analogies may suggest the way of creating artificial 
intelligence (Yukalov and Sornette 2009c).  

The processes of self-organization and decision making can be treated from
two different points of view, complementing each other. First, it is possible
to analyze the actual process of the appearance of structures in a complex 
system consisting of many agents that are characterized by their typical 
features and by their interactions with each other as well as with external 
fields and perturbations. This requires to consider the dynamical equations 
of such multi-agent statistical systems, whether this is a thermodynamic, 
biological, social system, or a neuron network in brain.  

The second part is the choice of the best way of presenting the results of 
solving the complicated dynamical systems, allowing for a convenient description 
and classification of the found solutions. This final stage is necessary for
a clear understanding of the obtained results and for their correct 
interpretation. Our goal in the present article has been exactly this descriptive 
stage.       

Our main aim here has been to show that the processes of self-organization in 
complex systems and of decision making by alive beings can be represented in 
the same mathematical language of the search for the highest probability 
corresponding to the most stable state or to the most preferable prospect. 
Several examples of complex systems that we have mentioned are already
well-known, and we have used them to illustrate that all of them can be described 
in the same probabilistic framework. The principal novelty of the present article 
is the development of a general probabilistic approach allowing us to describe 
self-organization and decision making in the same mathematical terms, thus
demonstrating that these two processes can be interpreted as been identical.

\vskip 5mm

{\bf Acknowledgment}

\vskip 3mm

The authors acknowledge financial support from the Swiss National Science
Foundation. We are grateful to M. Favre and E.P. Yukalova for many useful 
discussions and advice.

\newpage

{\Large{\bf References}}

\vskip 5mm

{\parindent=0pt

\vskip 2mm
Anderson, S.P., De Palma, A., and Thisse, J.F. 1992  
{\it Discrete Choice Theory of Product Differentiation}. 
Cambridge: Massachusetts Institute of Technology.

\vskip 2mm
Aquinas, T. 1265-1274
{\it Summa Theologica}, 
translation 1948 by Fathers of the English Dominican Province, Vol. 1.
New York: Benziger.

\vskip 2mm
Arecchi, F.T., Boccaletti, S., and Ramazza, P.L. 1999
Pattern formation and competition in nonlinear optics.
{\it Physics Reports} {\bf 318}, 1--83.

\vskip 2mm
Arias, A., Lazaro, E., Escarmis, C., and Domingo, E. 2001 
Molecular intermediates of fitness gain of an RNA virus: characterization 
of a mutant spectrum by biological and molecular cloning. 
{\it Journal of General Virology} {\bf 82}, 1049--1060.

\vskip 2mm
Ashby, W.R. 1947
Principles of the self-organizing dynamic system. 
{\it Journal of General Psychology} {\bf 37}, 125--128.

\vskip 2mm
Bansmann, J. et al. 2005
Magnetic and structural properties of isolated and assembled clusters.
{\it Surface Science Reports} {\bf 56}, 185--275.  

\vskip 2mm
Baumg\"{a}rnter, S. 2004
Thermodynamic models. In {\it Modelling in Ecological Economics}
(eds. J. Proops and P. Safonov), p. 102--129.
Cheltenham: Edward Elgar.

\vskip 2mm
Bernoulli, D. 1738
Exposition of a new theory on the measurement of risk.
{\it Proceedings of the Imperial Academy of Sciences of St. Petersburg}
{\bf 5}, 175--192.

\vskip 2mm
Berns, G.S., Laibson, D., and Loewenstein, G. 2007
Intertemporal choice: toward an integrative framework.
{\it Trends in Cognitive Sciences} {\bf 11}, 482--488.

\vskip 2mm
Birman, J.L., Nazmitdinov, R.G., and Yukalov, V.I. 2013
Effects of symmetry breaking in finite quantum systems.
{\it Physics Reports} {\bf 526}, 1--91. 

\vskip 2mm
Bohr, N. 1933
Light and life. 
{\it Nature} {\bf 131}, 421--423, 457--459.

\vskip 2mm
Bohr, N. 1958
{\it Atomic Physics and Human Knowledge}. New York: Wiley.

\vskip 2mm
Bouchaud, J.P. and Georges, A. 1990
Anomalous diffusion in disordered media: Statistical mechanisms, 
models and physical application.
{\it Physics Reports} {\bf 195}, 127--293.

\vskip 2mm
Brock, W.A. and Hommes, C.H. 1997 
A rational route to randomness.
{\it Econometrica} {\bf 65}, 1059--1095.

\vskip 2mm
Castellano, C., Fortunato, S., and Loreto, V. 2009
Statistical physics of social dynamics.
{\it Reviews of Modern Physics} {\bf 81}, 591--646. 

\vskip 2mm
Chialvo, D.R. 2010
Emergent complex neural dynamics.
{\it Nature Physics} {\bf 6}, 744--750.

\vskip 2mm
Chowdury, D. and Stauffer, D. 1999
A generalized spin model of financial markets.
{\it European Physical Journal B} {\bf 8}, 477--482. 

\vskip 2mm
Coleman, A.J. and Yukalov, V.I. 2000
{\it Reduced Density Matrices}. Berlin: Springer.

\vskip 2mm
Cowperthwaite, M.C., Bull, J.J., and Meyers, L.A. 2005 
Distributions of beneficial fitness effects in RNA. 
{\it Genetics} {\bf 170}, 1449--1457.

\vskip 2mm
Crozier, R.H. and Pamilo, P. 1979 
Frequency-dependent models for X-linked loci in kangaroos.
{\it Australian Journal of Biological Sciences} {\bf 32}, 469--474.

\vskip 2mm
Danilov, V.I. and Lambert-Mogiliansky, A. 2008
Measurable systems and behavioral sciences.
{\it Mathematical Social Sciences} {\bf 55}, 315--340.

\vskip 2mm
Danilov, V.I. and Lambert-Mogiliansky, A. 2011
Expected utility theory under non-classical uncertainty.
{\it Theory and Decision} {\bf 68}, 25--47.

\vskip 2mm
Dixit, A. and Besley, T. 1997
James Mirrlees contribution to the theory of information and incentives.
{\it Scandinavian Journal of Economics} {\bf 99}, 207--235.

\vskip 2mm
Dunbar, R. 2006
We believe.
{it New Scientist} {\bf 28}, 30--33.

\vskip 2mm
Edwards, A.W.F. 1972 
{\it Likelihood}. Cambridge: Cambridge University. 

\vskip 2mm
Encinas-Sanz, F., Leyva, I, and Guerra, J.M. 2000
Time-resolved spatiotemporal dynamics in a broad-area CO$_2$ laser.
{\it Physical Review A} {\bf 62}, 043821.

\vskip 2mm
Epstein, L.G. 2008
Living with risk.
{\it Review of Economic Studies} {\bf 75}, 1121--1141.

\vskip 2mm
Fingelkurts, A.A., Fingelkurts, A.A., and Neves, C.F.H. 2012
Machine consciousness and artificial thought: an operational 
architectonics model guided approach.
{\it Brain Research} {\bf 1428}, 80--92.

\vskip 2mm
Foley, D.K. 1994
A statistical equilibrium theory of markets.
{\it Journal of Economic Theory} {\bf 62}, 321--345. 

\vskip 2mm
Foley, D.K. 1996
Statistical equilibrium in a simple labor market.
{\it Metroeconomica} {\bf 47}, 125--147.  

\vskip 2mm
Florian, R. and Galam, S. 2000
Optimizing conflicts in the formation of strategic alliances.
{\it European Physical Journal B} {\bf 16}, 189--194. 

\vskip 2mm
Frederick, S., Loewenstien, G., and O'Donoghue, T. 2002
Time discounting and time preference: a critical review.
{\it Journal of Economic Literature} {\bf 40}, 351--401. 

\vskip 2mm
Galam, S. 1996
Fragmentation versus stability in bimodal coalitions.
{\it Physica A} {\bf 230}, 174--188.

\vskip 2mm
Galam, S. and Moscovici, S. 1991
Towards a theory of collective phenomena: consensus and attitude changes 
in groups.
{\it European Journal of Social Psychology} {\bf 21}, 49--74.

\vskip 2mm
Galam, S. and Walliser, B. 2010
Ising model versus normal form game.
{\it Physica A} {\bf 389}, 481--489.

\vskip 2mm
Galam, S. 2012
{\it Sociophysics: A Physicist's Psycho-Political Modeling of Phenomena}. 
New York: Springer.

\vskip 2mm
Gallo, I., Barra, A. and Contucci, P. 2009
Parameter evaluation of a simple mean-field model of social interactions.
{\it Mathematical Models and Methods in Applied Sciences} 
{\bf 19}, 1427--1439. 

\vskip 2mm
Gaudefroy, L. et al. 2009
Shell erosion and shape coexistence in $_{16}^{43}S_{27}$.
{\it Physical Review Letters} {\bf 102}, 092501.

\vskip 2mm
Gibbs, J.W. 1902
{\it Elementary Principles in Statistical Mechanics}. 
Oxford: Oxford University.

\vskip 2mm
Gibbs, J.W. 1928  
{\it Collected Works}. New York: Longmans, Vol. 1.

\vskip 2mm
J.W. Gibbs, J.W. 1931  
{\it Collected Works}. New York: Longmans, Vol. 2.

\vskip 2mm
Glansdorff, P. and Prigogine, I. 1971
{\it Thermodynamic Theory of Structure, Stability and Fluctuations}.
London: Wiley.

\vskip 2mm
Griffin, D. and Tversky, A. 1992
The weighing of evidence and the determinants of confidence.
{\it Cognitive Psychology} {\bf 24}, 411--435.

\vskip 2mm
Haken, H. 1983
{\it Advanced Synergetics}. Berlin: Springer.

\vskip 2mm
Haken, H. 2005 
Synergetics. In {\it Encyclopedia of Nonlinear Science} (ed. A. Scott), 
pp. 910--914. New York: Routledge.

\vskip 2mm
Harras, G. and Sornette, D. 2011
How to grow a bubble: A model of myopic adapting agents.
{\it Journal of Economic Behavior and Organization} {\bf 80} (1), 137--152.

\vskip 2mm
Heylighen, F. 2009 
Complexity and self-organization. In {\it Encyclopedia of Library and 
Information Sciences} (ed. M.J. Bates and M.N. Maack.). 
London: Taylor and Francis.

\vskip 2mm
Hooker, C.A., Gabbay, D.M., Thagard P. and Woods J. (eds.) 2011
{\it Philosophy of Complex Systems. 
Handbook of the Philosophy of Science}, Vol. 10.
Amsterdam: North Holland.

\vskip 2mm
Jaynes., E.T. 1957
Information theory and statistical mechanics.
{\it Physical Review} {\bf 106}, 620--630.

\vskip 2mm
Jona-Lasinio, G. 2001
Renormalization group and probability theory.
{\it Physics Reports} {\bf 352}, 439--458.

\vskip 2mm
Kant, I. 2007
{\it Critique of Judgement}. 
Oxford: Oxford University Press.

\vskip 2mm
Kauffman, S. 1996
{\it At Home in the Universe: 
The Search for the Laws of Self-Organization and Complexity}.
Oxford: Oxford University Press.

\vskip 2mm
Khinchin, A.I. 1949
{\it Mathematical Foundations of Statistical Mechanics}.
New York: Dover. 

\vskip 2mm
Klein, O. 1931
Zur Quantenmechanischen Begr\"{u}ndung des zweiten Hauptsatzes
der W\"{a}rmelehre. 
{\it Zeitschrift f\"{u}r Physik} {\bf 72}, 767--775.

\vskip 2mm
Krugman, P. 1996
{\it The Self-Organizing Economy}, 
Oxford: Blackwell.

\vskip 2mm
Kuehn, C. 2011
A mathematical framework for critical transitions: bifurcations,
fast-slow dynamics and stochastic dynamics.
{\it Physica D} {\bf 240}, 1020--1035.

\vskip 2mm
Kullback, S. and Leibler, R.A. 1951 
On information and sufficiency. 
{\it Annals of Mathematical Statistics} {\bf 22}, 79--86.

\vskip 2mm
Kullback, S. 1959 
{\it Information Theory and Statistics}. 
New York: Wiley.

\vskip 2mm
Kusmartsev, E.V. 2011
Statistical mechanics of economics.
{\it Physics Letters A} {\bf 375}, 966--973.

\vskip 2mm
Labini, F.S., Vasilyev, N.L., Pietronero, L. and Baryshev, Y. 2009
Absence of self-averaging and of homogeneity in the large scale galaxy 
distribution.
{\it Europhys. Lett.} {\bf 86}, 49001.

\vskip 2mm
Lambert-Mogiliansky, A., Zamir, S. and Zwirn, H. 2009
Type indeterminacy: a model of the Kahneman-Tversky man.
{\it Journal of Mathematical Psychology} {\bf 53}, 349--361. 

\vskip 2mm
Landau, L.D. and Lifshitz, E.M. 1980 
{\it Statistical Physics}.
Oxford: Butterworth-Heinemann.

\vskip 2mm
Lehrer, E. and Shmaya, E. 2006
A qualitative approach to quantum probability.
{\it Proceedings of Royal Society A} {\bf 462}, 2331--2344. 

\vskip 2mm
Leyva, I. and Guerra, J.M. 2002
Time-resolved pattern evolution in a large-aperture class A laser.
{\it Physical Review A} {\bf 66}, 023820.

\vskip 2mm
Loewenstein, G. and Thaler, R.H. 1989
Intertemporal choice.
{\it Journal of Economic Perspectives} {\bf 34}, 181--193.

\vskip 2mm
Luce, R.D. 1958
A probabilistic theory of utility.
{\it Econometrica} {\bf 26}, 193--224.

\vskip 2mm
Machina, M.J. 2008
Non-expected utility theory. 
In {\it New Palgrave Dictionary of Economics} 
(eds. S.N. Durlauf and L.E. Blume). New York: Macmillan.

\vskip 2mm
Mainzer, K. 2007
{\it Thinking in Complexity}. Berlin: Springer.

\vskip 2mm
Manly, B.F. 1976
Some examples of double exponential fitness functions.
{\it Heredity} {\bf 36}, 229--234.

\vskip 2mm
Marsili, M., Raffaelli, G. and Ponsot, B. 2009
Dynamic instability in generic model of multi-assets markets
{\it Journal of Economic Dynamics and Control} {\bf 33}, 1170--1181.

\vskip 2mm
Martin, G. and Lenormand T. 2008 
The distribution of beneficial and fixed mutation fitness effects 
close to an optimum. 
{\it Genetics} {\bf 179}, 907--916.

\vskip 2mm
Maturana, H. 1988
Reality: the search for objectivity or the quest for a compelling argument.
{\it Irish Journal of Psychology} {\bf 9}, 25--82.

\vskip 2mm
McCauley, J.L. 2003
Thermodynamic analogies in economics and finance: instability of markets.
{\it Physica A} {\bf 329}, 199--212.

\vskip 2mm
McFadden, D. 1974 
Conditional logit analysis of qualitative choice behavior. 
In {\it Frontiers in Econometrics} (ed. P. Zarembka), p. 105--142.
New York: Academic.

\vskip 2mm
McPherson, T. 1972
{\it The Argument From Design}. New York: Macmillan.

\vskip 2mm
M\'ezard, M., Parisi, G. and Virasoro, M. 1986
{\it Spin Glass Theory And Beyond: An Introduction To The Replica Method 
And Its Applications}.
Singapore: World Scientific.

\vskip 2mm
Napel, S. 2003
Aspiration adaptation in the ultimatum minigame.
{\it Games and Economic Behavior} {\bf 43}, 86--106.  

\vskip 2mm
Nichols, S. 2011
Experimental philosophy and the problem of free will.
{\it Science} {\bf 331}, 1401--1403.

\vskip 2mm
Nicolis, J.S. 1986
{\it Dynamics of Hierarchical Systems}. Berlin: Springer.

\vskip 2mm
Paley, W. 1802
{\it Natural Theology, or Evidences of the Existence and Attributes of 
the Deity Collected from the Appearances of Nature}
London: Faulder.

\vskip 2mm
Pask, G. 1996
Heinz von Foerster's self-organization, the progenitor of conversation 
and interaction theories.
{\it Systems Research} {\bf 13}, 349--362. 

\vskip 2mm
Phothos, E.M. and Busemeyer, J.R. 2009
A quantum probability model explanation for violations of rational 
decision making.
{\it Proceedings of Royal Society B} {\bf 276}, 2171--2178. 

\vskip 2mm
Poerksen, B. 2003
At each and every moment I can decide who I am.
{\it Cybernetics and Human Knowing} {\bf 10}, 9--26.

\vskip 2mm
Rambaud, S.C. and Torrecillas, M.J. 2005
Some considerations on the social discount rate.
{\it Environmental Science and Policy} {\bf 8}, 343--355. 

\vskip 2mm
Russell, D.A. 1996 
The significance of the extinction of the dinosaurs.
{\it Geological Society of America} {\bf 307}, 381--388.

\vskip 2mm
Saakian, D.B., Martirosyan, A.S. and Hu, C.K. 2010
Different fitnesses for in vivo and in vitro evolutions due to the 
finite generation-time effect
{\it Physical Review E} {\bf 81}, 061913.

\vskip 2mm
Samuelson, P. 1937
A note on measurement of utility.
{\it Review of Economic Studies} {\bf 4}, 155--161.

\vskip 2mm
Savage, L.J. 1954 
{\it The Foundations of Statistics}. New York: Wiley.

\vskip 2mm
Schelling, T.C. 1978
{\it Micromotives and Macrobehavior}. New York: Norton.

\vskip 2mm
Selten, R. 1998
Aspiration adaptation theory.
{\it Journal of Mathematical Psychology} {\bf 42}, 191--214.

\vskip 2mm
Shafir, E.B., Smith, E.E. and Osherson, D.N. 1990
Typicality and reasoning fallacy.
{\it Memory and Cognition} {\bf 18}, 229--239.

\vskip 2mm
Shannon, C.E. and Weaver, W. 1949
{\it  Mathematical Theory of Communication}. 
Urban: University of Illinois.

\vskip 2mm
Shumovsky, A.S. and Yukalov, V.I. 1982
Exact solutions for heterophase ferromagnets.
{\it Physica A} {\bf 110}, 518--534.

\vskip 2mm
Simon, H.A. 1959
Theories of decision making in economics and behavioral science.
{\it American Economic Review} {\bf 49}, 253--283. 

\vskip 2mm
Smith, E. and Foley, D.K. 2008
Classical thermodynamics and economic general equilibrium theory.
{\it Journal of Economic Dynamics and Control} {\bf 32}, 7--65.

\vskip 2mm
Sornette, D. 2003
{\it Why Stock Markets Crash}.
Princeton: Princeton University.

\vskip 2mm
Sornette, D. 2006
{\it Critical Phenomena in Natural Sciences}.
Berlin: Springer.

\vskip 2mm
Stauffer, D. 2008
Social applications of two-dimensional Ising models.
{\it American Journal of Physics} {\bf 76}, 470--473.

\vskip 2mm
Turalska, M., West, B., and Grigolini, P. 2011
Temporal complexity of the order parameter at the phase transition. 
{\it Physical Review E} {\bf 83}, 061142.

\vskip 2mm
Tversky, A. and Kahneman, D. 1973
Availability: a heuristic for judging frequency and probability,
{\it Cognitive Psychology} {\bf 5}, 207--232.

\vskip 2mm
Tversky, A. and Kahneman, D. 1980
Judgements of and by representativeness,
In {\it Judgements Under Uncertainty: Heuristics and Biases}
(eds. D. Kahneman, P. Slovic and A. Tversky), pp. 84--98. 
New York: Cambridge University.

\vskip 2mm
Tversky, A. and Kahneman, D. 1983
Extensional versus intuitive reasoning: the conjunction fallacy in 
probability judgment.
{\it Psychological Review} {\bf 90}, 293--315. 

\vskip 2mm
Vanni, F., Lukovic, M., and Grigolini P. 2011
Criticality and transmission of information in a swarm of cooperative units.
{\it Physical Review Letters} {\bf 107}, 078103.

\vskip 2mm
von Delft, J. 2001
Superconductivity in ultrasmall metallic grains.
{\it Annalen der Physik} {\bf 10}, 1--60.

\vskip 2mm
von Foerster, H. 1960
On self-organising systems and their environments, 
In {\it Self-Organising Systems} (eds. M.C. Yovits and S. Cameron), pp. 30--50. 
London: Pergamon.

\vskip 2mm
von Foerster, H. and Pask, G. 1960
A predictive evolutionary model.
{\it Cybernetica} {\bf 4}, 258--300.

\vskip 2mm
von Foerster, H. 1995
{\it Cybernetics of Cybernetics}.
Minneapolis: Future Systems.

\vskip 2mm
von Foerster, H. 1999
Ethics and second-order cybernetics.
{\it Stanford Humanity Review} {\bf 4}, 308--319.

\vskip 2mm
von Foerster, H. 2003
Action without utility: an immodest proposal for the cognitive foundation 
of behavior. {\it Cybernetics and Human Knowing} {\bf 10}, 27--50.

\vskip 2mm
von Galsserfeld, E. 1991
Distinguishing the observer: an attempt at interpreting Maturana.
{\it Methodologia} {\bf 5}, 92--111.

\vskip 2mm
von Glasserfeld, E. 1996
Cybernetics and the art of living.
{\it Cybernetics and Systems} {\bf 27}, 489--497.

\vskip 2mm
von Neumann, J. 1955 
{\it Mathematical Foundations of Quantum Mechanics}.
Princeton: Princeton University.

\vskip 2mm
von Neumann, J. and  Morgenstern, O. 1953 
{\it Theory of Games and Economic Behavior}.
Princeton: Princeton University.

\vskip 2mm
Werner, G. 2012
From brain states to mental phenomena via phase space transitions and
renormalization group transformation: proposal of a theory.
{\it Cognitive Neurodynamics} {\bf 6}, 199--202.

\vskip 2mm
West, B.J. and Grigolini, P. 2010
A psychological model of decision making.
{\it Physica A} {\bf 389}, 3580--3587. 

\vskip 2mm
Wiener, N. 1961
{\it Cybernetics: Or Control and Communication in the Animal and the Machine}. 
Paris: Hermann and Cie.

\vskip 2mm
Wissner-Gross, A.D. and Freer, C.E. 2013
Causal entropic forces.
{\it Physical Review Letters} {\bf 110}, 168702. 

\vskip 2mm
Yakovenko, V.M. and Rosser, J.B. 2009
Statistical mechanics of money, wealth, and income.
{\it Reviews of Modern Physics} {\bf 81}, 1703--1725.

\vskip 2mm
Yates, J.F. and Carlson, B.W. 1986
Conjunction errors: evidence for multiple judgment procedures, 
including signed summation.
{\it Organizational Behavior and Human Decision Processes} {\bf 37}, 230--253. 

\vskip 2mm
Yukalov V.I. 1981
A new method in the theory of phase transitions.
{\it Physics Letters A} {\bf 81}, 249--251.

\vskip 2mm
Yukalov, V.I. 1991
Phase transitions and heterophase fluctuations.
{\it Physics Reports} {\bf 208}, 395--492.

\vskip 2mm
Yukalov, V.I. 2000
Turbulent photon filamentation in resonant media.
{\it Physics Letters A} {\bf 278}, 30--34.

\vskip 2mm
Yukalov, V.I. 2001a
Principle of pattern selection for nonequilibrium phenomena.
{\it Physics Letters A} {\bf 284}, 91--98. 

\vskip 2mm
Yukalov, V.I. 2001b
Probabilistic approach to pattern selection.
{\it Physica A} {\bf 291}, 255--274.

\vskip 2mm
Yukalov, V.I. 2001c
Self-similar approach to market analysis.
{\it European Physical Journal B} {\bf 20}, 609--617.

\vskip 2mm
Yukalov, V.I. 2002
Matrix order indices in statistical mechanics.
{\it Physica A} {\bf 310}, 413--434.

\vskip 2mm
Yukalov, V.I. 2003a
Mesoscopic phase fluctuations: general phenomenon in condensed mater.
{\it International Journal of Modern Physics B}, {\bf 17}, 2333--2358. 

\vskip 2mm
Yukalov, V.I. 2003b
Expansion exponents for nonequilibrium systems.
{\it Physica A} {\bf 320}, 149--168.

\vskip 2mm
Yukalov, V.I. 2007
Representative ensembles in statistical mechanics.
{\it International Journal of Modern Physics B} {\bf 21}, 69--86.

\vskip 2mm
Yukalov, V.I. 2011
Nonequilibrium representative ensembles for isolated quantum system.
{\it Physics Letters A} {\bf 375}, 2797--2801.

\vskip 2mm
Yukalov, V.I. and Shumovsky, A.S. 1990
{\it Lectures on Phase Transitions}.
Singapore: World Scientific. 

\vskip 2mm
Yukalov, V.I. and Sornette, D. 2008
Quantum decision theory as quantum theory of measurement. 
{\it Physics Letters A} {\bf 372}, 6867--6871.

\vskip 2mm
Yukalov, V.I. and Sornette, D. 2009a
Physics of risk and uncertainty in quantum decision making. 
{\it European Physical Journal B} {\bf 71}, 533--548.

\vskip 2mm
Yukalov, V.I. and Sornette, D. 2009b
Processing information in quantum decision theory. 
{\it Entropy} {\bf 11}, 1073--1120.

\vskip 2mm
Yukalov, V.I. and Sornette, D. 2009c
Scheme of thinking quantum systems. 
{\it Laser Physics Letters} {\bf 6}, 833--839.

\vskip 2mm
Yukalov, V.I. and Sornette, D. 2010
Mathematical structure of quantum decision theory. 
{\it Advances in Complex Systems} {\bf 13}, 659--698.

\vskip 2mm
Yukalov, V.I. and Sornette, D. 2011
Decision theory with prospect interference and entanglement. 
{\it Theory and Decision} {\bf 70}, 283--328.

\vskip 2mm
Zhou, W.X. and Sornette, D. 2007
Self-organizing Ising model of financial markets.
{\it European Physical Journal B} {\bf 55}, 175--181.

\vskip 2mm
Zurek, W.H. 2003
Decoherence, selection, and the quantum origins of the classical. 
{\it Reviews of Modern Physics} {\bf 75}, 715--775. 

}

\end{document}